\documentclass[11pt,graphicx,subfigure,axodraw]{article}
\usepackage{amsfonts}
\usepackage{amssymb}
\usepackage{amsmath,amsfonts,amssymb}
\usepackage{epstopdf}
\usepackage[section]{placeins}
\usepackage{hyperref}
\usepackage{booktabs}
\usepackage{cleveref}
\crefname{equation}{Eq.}{Eqs.}
\crefname{figure}{Fig.}{Figs.}
\crefname{table}{Table}{Tables}
\crefname{section}{Section}{Sections}

\usepackage[usenames,dvipsnames]{color}
\usepackage{graphicx}
\usepackage{subfigure}
\usepackage{float}
\def\rmuu{\gamma^{\mu}}
\def\rmud{\gamma_{\mu}}
\def\PL{{1-\gamma_5\over 2}}
\def\PR{{1+\gamma_5\over 2}}
\def\sinW2{\sin^2\theta_W}
\def\AEM{\alpha_{EM}}
\def\mul{M_{\tilde{u} L}^2}
\def\mur{M_{\tilde{u} R}^2}
\def\mdl{M_{\tilde{d} L}^2}
\def\mdr{M_{\tilde{d} R}^2}
\def\mz2{M_{z}^2}
\def\c2b{\cos 2\beta}
\def\au{A_u}
\def\ad{A_d}
\def\cob{\cot \beta}
\def\v#1{v_#1}
\def\tb{\tan\beta}
\def\epem{$e^+e^-$}
\def\KK{$K^0$-$\overline{K^0}$}
\def\wi{\omega_i}
\def\xj{\chi_j}
\def\Wmu{W_\mu}
\def\Wnu{W_\nu}
\def\m#1{{\tilde m}_#1}
\def\mH{m_H}
\def\mw#1{{\tilde m}_{\omega #1}}
\def\mx#1{{\tilde m}_{\chi^{0}_#1}}
\def\mc#1{{\tilde m}_{\chi^{+}_#1}}
\def\mwi{{\tilde m}_{\omega i}}
\def\mxi{{\tilde m}_{\chi^{0}_i}}
\def\mci{{\tilde m}_{\chi^{+}_i}}

\def\ch{{\tilde\chi^{+}_1}}
\def\c2{{\tilde\chi^{+}_2}}

\def\tt{{\tilde\theta}}

\def\tp{{\tilde\phi}}

\def\mz{M_z}
\def\sw{\sin\theta_W}
\def\cw{\cos\theta_W}
\def\cb{\cos\beta}
\def\sb{\sin\beta}
\def\rwi{r_{\omega i}}
\def\rxj{r_{\chi j}}
\def\rfp{r_f'}
\def\Kik{K_{ik}}
\def\Fq2{F_{2}(q^2)}
\def\f{\({\cal F}\)}
\def\d1{{\f(\tilde c;\tilde s;\tilde W)+ \f(\tilde c;\tilde \mu;\tilde W)}}
\def\tw{\tan\theta_W}
\def\sec2w{sec^2\theta_W}
\begin{document}
\baselineskip 18pt
\def\today{\ifcase\month\or
 January\or February\or March\or April\or May\or June\or
 July\or August\or September\or October\or November\or December\fi
 \space\number\day, \number\year}
\def\thebibliography#1{\section*{References\markboth
 {References}{References}}\list
 {[\arabic{enumi}]}{\settowidth\labelwidth{[#1]}
 \leftmargin\labelwidth
 \advance\leftmargin\labelsep
 \usecounter{enumi}}
 \def\newblock{\hskip .11em plus .33em minus .07em}
 \sloppy
 \sfcode`\.=1000\relax}
\let\endthebibliography=\endlist
\def\lsim{\ ^<\llap{$_\sim$}\ }
\def\gsim{\ ^>\llap{$_\sim$}\ }
\def\r2{\sqrt 2}
\def\beq{\begin{equation}}
\def\eeq{\end{equation}}
\def\beqn{\begin{eqnarray}}
\def\eeqn{\end{eqnarray}}
\def\rmuu{\gamma^{\mu}}
\def\rmud{\gamma_{\mu}}
\def\PL{{1-\gamma_5\over 2}}
\def\PR{{1+\gamma_5\over 2}}
\def\sinW2{\sin^2\theta_W}
\def\AEM{\alpha_{EM}}
\def\mul{M_{\tilde{u} L}^2}
\def\mur{M_{\tilde{u} R}^2}
\def\mdl{M_{\tilde{d} L}^2}
\def\mdr{M_{\tilde{d} R}^2}
\def\mz2{M_{z}^2}
\def\c2b{\cos 2\beta}
\def\au{A_u}
\def\ad{A_d}
\def\cob{\cot \beta}
\def\v#1{v_#1}
\def\tb{\tan\beta}
\def\epem{$e^+e^-$}
\def\KK{$K^0$-$\bar{K^0}$}
\def\wi{\omega_i}
\def\xj{\chi_j}
\def\Wmu{W_\mu}
\def\Wnu{W_\nu}
\def\m#1{{\tilde m}_#1}
\def\mH{m_H}
\def\mw#1{{\tilde m}_{\omega #1}}
\def\mx#1{{\tilde m}_{\chi^{0}_#1}}
\def\mc#1{{\tilde m}_{\chi^{+}_#1}}
\def\mwi{{\tilde m}_{\omega i}}
\def\mxi{{\tilde m}_{\chi^{0}_i}}
\def\mci{{\tilde m}_{\chi^{+}_i}}
\def\mz{M_z}
\def\sw{\sin\theta_W}
\def\cw{\cos\theta_W}
\def\cb{\cos\beta}
\def\sb{\sin\beta}
\def\rwi{r_{\omega i}}
\def\rxj{r_{\chi j}}
\def\rfp{r_f'}
\def\Kik{K_{ik}}
\def\Fq2{F_{2}(q^2)}
\def\f{\({\cal F}\)}
\def\d1{{\f(\tilde c;\tilde s;\tilde W)+ \f(\tilde c;\tilde \mu;\tilde W)}}
\def\tw{\tan\theta_W}
\def\sec2w{sec^2\theta_W}
\def\ch{{\tilde\chi^{+}_1}}
\def\c2{{\tilde\chi^{+}_2}}

\def\tt{{\tilde\theta}}

\def\tp{{\tilde\phi}}

\def\mz{M_z}
\def\sw{\sin\theta_W}
\def\cw{\cos\theta_W}
\def\cb{\cos\beta}
\def\sb{\sin\beta}
\def\rwi{r_{\omega i}}
\def\rxj{r_{\chi j}}
\def\rfp{r_f'}
\def\Kik{K_{ik}}
\def\Fq2{F_{2}(q^2)}
\def\f{\({\cal F}\)}
\def\d1{{\f(\tilde c;\tilde s;\tilde W)+ \f(\tilde c;\tilde \mu;\tilde W)}}

\def\b{$\cal{B}(\tau\to\mu \gamma)$~}


\def\tw{\tan\theta_W}
\def\sec2w{sec^2\theta_W}
\newcommand{\pn}[1]{{\color{red}{#1}}}

\begin{titlepage}
\begin{center}
{\large {\bf
Higgs Mass  and  CP violating Phases Implications on the SUSY Breaking Scale in MSSM}}\\


\vskip 0.5 true cm
 Tarek  Ibrahim$^{a}$\footnote{Email:
tibrahim@zewailcity.edu.eg} and Anas Zorik$^{b}$\footnote{Email: zorik@alexu.edu.eg}
\vskip 0.5 true cm
\end{center}

\noindent
{$^{a}$ Center of Fundamental Physics CFP,
 Zewail City of Science and Technology, 6th October City, Giza 12578,  Egypt. }\\
{$^{b}$ 
 Department of  Physics, Faculty of Science,
University of Alexandria, Alexandria 21511, Egypt. } \\

\vskip 0.5 true cm

\centerline{\bf Abstract}
 The large Higgs mass  $m_h \approx 125 GeV$ can raise the values of the SUSY breaking parameters in the Minimal Supersymmetric Standard Model MSSM to the range of several TeV scale.
The CP violating phases in the MSSM can induce EDMs of the fermions in the theory already in conflict with the current experimental upper limit unless the masses of the SUSY partners are on the heavy side.
 We invistigate the implications of both constraints, the Higgs Mass  and the complete set of SUSY CP violating phases in the analysis of the electric dipole moments EDMs on the SUSY breaking parameters and thus on the SUSY mass spectrum.
We use the electric dipole moments  of the electron, the neutron and the proton as our probes in this study.

 \noindent
{\scriptsize
}

\medskip

\end{titlepage}
\section{Introduction \label{sec1}}
In the minimual Supersymmetric Standard Model (MSSM) the loop corrections to the effective potential make very important contributions to the Higgs masses. Thus in the absence of the loop corrections the lightest Higgs mass satisfies the inequality $m_h < M_Z$ already in contradiction with the 125 GeV measured value.
However, with the inclusion of radiative corrections the lightest Higgs mass can be lifted above $M_Z$ and can reach the measured value. The major corrections to the lightest Higgs mass come from the stop exchange contribution and the analysis can be exteded to include the sbottom exchange which is much smaller. In this paper we only consider the stop corrections to the Higgs mass spectrum ~\cite{Coleman}-\cite{Ibrahim:11}.

The search for SUSY partners at colliders and the mass of the Higgs boson push the limits of the softly broken supersymmetry parameters up to the several TeV scale region. As is well known these SUSY parameters are generally complex and bring new sources of CP violation ~\cite{EDMS}, ~\cite{Ibrahimold}. The natural size of these phases is large, typically $O(1)$, and can induce electric dipole moments for the electron, the neutron and the proton. The current experimental limits on these moments are $|d_e|<4.1\times 10^{-30} e.cm, |d_n|< 1.8\times 10^{-26} e.cm, |d_p|<2.1\times 10^{-25} e.cm$ ~\cite{sm}.
The several TeV SUSY partners masses that are running in the EDM loops can suppress the EDMs within their upper limits.

Thus the discovery of the Higgs mass at 125 GeV ~\cite{Atlas1}-\cite{Ellis2} would introduce a kinematic suppression to the EDMs and can allow the CP violaing phases to be O(1). The analysis of the electric dipole moments of the electron and nucleons in MSSM with the most general allowed set of CP violating phases shows that the EDMs depend only on certain combinations of the CP phases.
We identify all of these seven combinations and their structure from the SUSY CP phases of the different complex parameters.
In this part of the analysis we include all one loop diagrams with the gluino, the chargino and the neutralino exchange for the electric dipole and the chromoelectric diople operators allowing for all phases. We also analyse the two loop diagrams which contribute to the purely gluonic dimension six operator.
We thus investigate the effects on the SUSY breakig scale of the discovered Higgs mass and the natural values of the CP violating phases constrained by the upper limits on the electron and nucleon EDMs.

The outline of the rest of the paper is as follows: In \cref{sec2} we give the Higgs mass corretions for the stop-top loops with the inclusion of CP violating phases.
In \cref{sec3} we discuss complete set of CP violating phases that arise from the EDMs analyses of the electron and the nucleons.
In \cref{sec4} we give a numerical analysis of the limits on the SUSY masses that can be 
calculated using the results of \cref{sec3}. Conclusions are given in \cref{sec5}.
Further details on the EDM components of quarks and electron are given 
 in \cref{sec6}.

\section{Higgs mass correction from quark-squark exchange  loops}\label{sec2}
In MSSM, ~\cite{Arnowitt, Pilaftsis, Higgs:1} the Higgs sector at the one loop level is described by the scalar potential
\begin{align}
V(H_1,H_2) = V_0 + \Delta V
\end{align}
The tree level potenital is given by
\begin{align}
V_0= m^2_1 |H_1|^2 + m^2_2 |H_2|^2 +(m^2_3 H_1.H_2 + h.c.)\\
+\frac{g^2_2 +g^2_1}{8} |H_1|^4 + \frac{g^2_2 +g^2_1}{8} |H_2|^4 -\frac{g^2_2}{2} |H_1.H_2|^2
+\frac{g^2_2 -g^2_1}{8} |H_1|^2 |H_2|^2
\end{align}

where $m^2_1=m^2_{H_1} +|\mu|^2$, $m^2_2=m^2_{H_2} +|\mu|^2$, $m^2_3=|\mu B|$ and $m_{H_{1,2}}$ and $B$ are the soft SUSY breaking parameters, and $\Delta V$ is the one loop correction to the effective potential and we consider here the largest contribution from the top-stop loops ~\cite{Coleman}
\begin{align}
\Delta V= \frac{3}{32 \pi^2}\{ \sum_{a}  m^4_{\tilde{t}_a} (\log{\frac{m^2_{\tilde{t_a}}}{Q^2}}-\frac{3}{2})
-2 m^4_t(\log{\frac{m^2_t}{Q^2}}-\frac{3}{2})\}
\end{align}
where  $a=1,2$ stands for the two eigen values of the squark $\tilde{t}$ masses.

As shown in Ref. ~\cite{Tarekhiggs}, the CP violating effects in the one loop effective potential induce CP violating phases to the Higgs VEVs through the minimization of the effective potential. One can parametrize this effect by the two phases $\chi_1$ and $\chi_2$ for the two Higgs where
\begin{align}
(H_1) = \left(\begin{matrix} H_1^0   \cr
 H_1^- \cr
\end{matrix}\right)=\frac{e^{i\chi_1}}{\sqrt{2}} \left(\begin{matrix} v_1+\phi_1+i\psi_1 \cr
 H_1^-\cr
\end{matrix}\right)\\
(H_2) = \left(\begin{matrix} H_2^+   \cr
 H_2^0 \cr
\end{matrix}\right)=\frac{e^{i\chi_2}}{\sqrt{2}} \left(\begin{matrix} H_2^+ \cr
 v_2+\phi_2+i\psi_2\cr
\end{matrix}\right)
\end{align}
The non-vanishing of the phases $\chi_1$ and $\chi_2$ can be seen by looking at the minimization of the effective potential.The variations with respect to the fields $\phi_1$, $\phi_2$, $\psi_1$ and $\psi_2$ gives the following conditions
\begin{align}
-\frac{1}{v_1} (\frac{\partial \Delta V}{\partial \phi_1})_0=m_1^2 +\frac{g_2^2+g_1^2}{8} (v_1^2 -v_2^2) + m_3^2 tan \beta \cos(\chi_1 +\chi_2)\\
-\frac{1}{v_2} (\frac{\partial \Delta V}{\partial \phi_2})_0=m_2^2 -\frac{g_2^2+g_1^2}{8} (v_1^2 -v_2^2) + m_3^2 cot \beta \cos(\chi_1 +\chi_2)\\
\frac{1}{v_1} (\frac{\partial \Delta V}{\partial \psi_2})_0=\frac{1}{v_2} (\frac{\partial \Delta V}{\partial \psi_1})_0=m_3^2 \sin(\chi_1+\chi_2)
\end{align}
where the subscript $0$ means that the quantities are evaluated at the point $\phi_1=\phi_2=\psi_1=\psi_2=0$.
With the inclusion of the stop-top contributions one finds that the phase $\chi_1 +\chi_2$ is determined by the equation
\begin{align}
m_3^2 \sin(\chi_1+\chi_2) = \frac{1}{2} \beta_{h_t} |\mu| |A_t| \sin\gamma_t f_1(m^2_{\tilde{t_1}}, m^2_{\tilde{t_2}})
\end{align}
where 
\begin{align}
\beta_{h_t}=\frac{3 h^2_t}{16 \pi^2}, ~ \gamma_t=\alpha_t +\theta_{\mu}
\end{align}

The phase $\alpha_t$ is the phase of the trilinear coupling $A_t$ of the flavor $t$ and $\theta_{\mu}$ is the phase of the bilinear coupling $\mu$ and $h_t$ is Yukawa coupling of the top quark.  The function $f(x,y)$ is given by
\begin{align}
f_1(x,y) = -2 +\log \frac{xy}{Q^4}+\frac{y+x}{y-x} \log \frac{y}{x}
\end{align}
It is clear that the CP violating phase $\chi_1+\chi_2$ vanishes when the CP violating phases are set to zero in the soft SUSY breaking parameters.

To construct the mass squared matrix of the Higgs scalars we need to calculate the quantities
\begin{align}
M^2_{ab}= (\frac{\partial^2 V}{\partial \Phi_a \partial \Phi_b})_0
\end{align}
where $\Phi_a (a=1-4)$ are defined by
\begin{align}
\{\Phi_a\} =\{\phi_1, \phi_2, \psi_1, \psi_2\}
\end{align}
and again the subscript $0$ means that we set $\phi_1=\phi_2=\psi_1=\psi_2=0$ after the computation of the Higgs mass matrix.
The matrix elements are coming from the tree level and from the loop conributions as
\begin{align}
M^2_{ab} = M^{2(0)}_{ab} +\Delta M^2_{ab}
\end{align}

Computation of the $4\times 4$ Higgs mass$^2$ matrix in the basis of $\{\phi_1, \phi_2, \psi_1, \psi_2\}$ gives
\begin{align}
M^2=
 \left(\begin{matrix} M^2_Z c^2_{\beta}+M^2_A s^2_{\beta}+\Delta_{11} &-( M^2_Z +M^2_A)s_{\beta} c_{\beta}+\Delta_{12}  &\Delta_{13} s_{\beta}  &\Delta_{13}c_{\beta}  \cr
 -( M^2_Z +M^2_A)s_{\beta} c_{\beta}+\Delta_{12}  &  M^2_Z s^2_{\beta}+M^2_A c^2_{\beta}+\Delta_{22} & \Delta_{23}s_{\beta}  &  \Delta_{23}c_{\beta} \cr 
 \Delta_{13}s_{\beta}  &   \Delta_{23}s_{\beta} & ( M^2_A +\Delta_{33})s^2_{\beta} & ( M^2_A +\Delta_{33})s_{\beta}c_{\beta}\cr 
\Delta_{13} c_{\beta} &  \Delta_{23} c_{\beta} & ( M^2_A +\Delta_{33})s_{\beta}c_{\beta} & ( M^2_A +\Delta_{33})c^2_{\beta} \cr
\end{matrix}\right)
\label{8}
\end{align}
where $c_{\beta} =\cos\beta, s_{\beta}=\sin\beta$. The explicit Q dependence has been absorbed in the term $M^2_A$ as follows
\begin{align}
M^2_A= \frac{1}{c_{\beta} s_{\beta}}\{-m^2_3 \cos(\chi_1+\chi_2) + \frac{1}{2} \beta_{h_t} |A_t| |\mu| \cos\gamma_t f_1(m^2_{\tilde{t_1}}, m^2_{\tilde{t_2}})\}
\end{align}
The first term is the tree term while the second  term comes from the stop  contributions.
The corrections $\Delta_{ij}$ are given by
\begin{align}
\Delta_{11 }=-2\beta_{h_t} m^2_t |\mu|^2 \frac{(|A_t|\cos\gamma_t-|\mu| cot\beta)^2}{(m^2_{\tilde{t_1}}-m^2_{\tilde{t_2}})^2} f_2(m^2_{\tilde{t_1}}, m^2_{\tilde{t_2}}),\\
\Delta_{22 }=-2\beta_{h_t} m^2_t |A_t|^2 \frac{(|A_t|-|\mu| cot\beta \cos\gamma_t)^2}{(m^2_{\tilde{t_1}}-m^2_{\tilde{t_2}})^2} f_2(m^2_{\tilde{t_1}}, m^2_{\tilde{t_2}})\nonumber\\
+2\beta_{h_t} m^2_t \ln(\frac{m^2_{\tilde{t_1}}m^2_{\tilde{t_2}}
}{m^4_t}) + 4\beta_{h_t} m^2_t |A_t| \frac{(|A_t|-|\mu| cot\beta \cos\gamma_t)}{(m^2_{\tilde{t_1}}-m^2_{\tilde{t_2}})}\ln(\frac{m^2_{\tilde{t_1}}}{m^2_{\tilde{t_2}}})   ,\\
\Delta_{12 }= -2\beta_{h_t} m^2_t |\mu| \frac{(|A_t|\cos\gamma_t-|\mu| cot\beta )}{(m^2_{\tilde{t_1}}-m^2_{\tilde{t_2}})}\ln(\frac{m^2_{\tilde{t_1}}}{m^2_{\tilde{t_2}}}) \nonumber\\
+2 \beta_{h_t}m^2_t |\mu||A_t| \frac{(|A_t| \cos\gamma_t-|\mu| cot\beta)(|A_t| -|\mu| cot\beta\cos\gamma_t)}{(m^2_{\tilde{t_1}}-m^2_{\tilde{t_2}})^2}  f_2(m^2_{\tilde{t_1}}, m^2_{\tilde{t_2}}),\\
\Delta_{13 }=-2 \beta_{h_t}m^2_t |\mu|^2|A_t| \sin\gamma_t \frac{(|\mu| cot\beta-|A_t| \cos\gamma_t)}{\sin\beta(m^2_{\tilde{t_1}}-m^2_{\tilde{t_2}})^2}  f_2(m^2_{\tilde{t_1}}, m^2_{\tilde{t_2}}) ,\\
\Delta_{23 }=-2 \beta_{h_t}m^2_t |\mu||A_t|^2 \sin\gamma_t \frac{(|A_t| -|\mu|cot\beta \cos\gamma_t)}{\sin\beta(m^2_{\tilde{t_1}}-m^2_{\tilde{t_2}})^2}  f_2(m^2_{\tilde{t_1}}, m^2_{\tilde{t_2}}) \nonumber\\
+ 2 \beta_{h_t} m^2_t |\mu| |A_t|\frac{\sin\gamma_t}{\sin\beta (m^2_{\tilde{t_1}}-m^2_{\tilde{t_2}})}\ln(\frac{m^2_{\tilde{t_1}}}{m^2_{\tilde{t_2}}}),\\
\Delta_{33 }=-2 \beta_{h_t}m^2_t |\mu|^2|A_t|^2 \sin^2\gamma_t \frac{1}{\sin^2\beta(m^2_{\tilde{t_1}}-m^2_{\tilde{t_2}})^2}  f_2(m^2_{\tilde{t_1}}, m^2_{\tilde{t_2}}) 
\end{align}

where $f_2(x,y)$ is given by
\begin{align}
f_2(x,y)=-2 + \frac{y+x}{y-x} \ln \frac{y}{x}
\end{align}
One may reduce the $4\times 4$ matrix of Eq.16 by introducing the new basis $\phi_1, \phi_2, \psi_{1D}, \psi_{2D}$ where
\begin{align}
\psi_{1D}= \sin\beta \psi_1 + \cos\beta \psi_2\\
\psi_{2D}= -\cos\beta \psi_1 +\sin\beta \psi_2
\end{align}
In this basis the field $\psi_{2D}$ decouples from the other fields with a vanishing mass and is the Goldstone field. The Higgs mass$^2$ matrix of the remaining three fields is given by
\begin{align}
M^2_{(Higgs)}=
 \left(\begin{matrix} M^2_Z c^2_{\beta}+M^2_A s^2_{\beta}+\Delta_{11} &-( M^2_Z +M^2_A)s_{\beta} c_{\beta}+\Delta_{12}  &\Delta_{13}    \cr
 -( M^2_Z +M^2_A)s_{\beta} c_{\beta}+\Delta_{12}  &  M^2_Z s^2_{\beta}+M^2_A c^2_{\beta}+\Delta_{22} & \Delta_{23}   \cr 
 \Delta_{13}  &   \Delta_{23} & ( M^2_A +\Delta_{33}) & \cr 
\end{matrix}\right)
\label{8}
\end{align}

\section{SUSY CP violating Phases  \label{sec3}}
\begin{figure}[t]
\begin{center}
{\rotatebox{0}{\resizebox*{10cm}{!}{\includegraphics{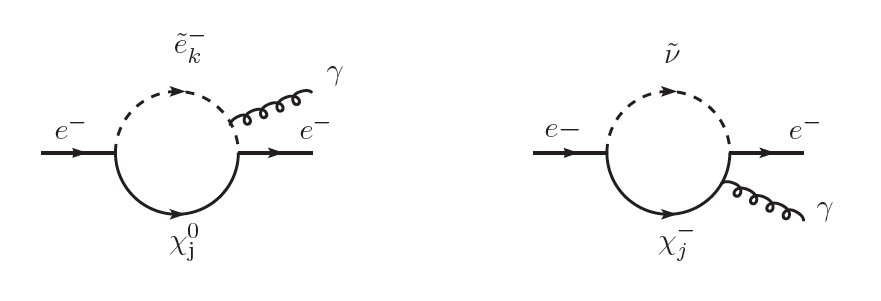}}\hglue5mm}}
\caption{The neutralino-slepton  exchange diagram (left) and the chargino -sneutrino exchange diagram (right) that contribute
to the  electric dipole moment of the electron in MSSM.}
 \label{fig1}
\end{center}
\end{figure}

In the analysis below we shall use the notation of Ref. ~\cite{Tarek3}. The loop contributions to the EDMs of quarks and leptons are calculated and we exhibit in \cref{fig1} the one loop contribution to the electron EDM from neutralino and chargino loops.

In MSSM the CP violating phases responsible for the quarks and leptons EDMs arise from the soft susy breaking part of the potential. These phases are $\theta_{\mu}$ of $\mu$, $\alpha_{f}$ of $A_f$,  the phases $\xi_i (i=1,2,3)$  of the three gaugino masses $\tilde{m}_i$ and $\chi_i (i=1,2)$ are the phases of the vacuum expectation values of the Higgs field as outlined in the previous section.

The chagino contribution to the EDM of the electron is given by
\begin{align}
d^{\chi^-}_{e}=\frac{\alpha_{EM}}{4\pi \sin^2 \theta} m^2_{\tilde{\nu_e}} \sum_{j=1,2} \tilde{m}_j Im (\kappa_e U^*_{j2}V^*_{j1} A(\frac{\tilde{m}^2_j}{m^2_{\tilde{\nu_e}}})
\end{align}
where $A(r) = \frac{2}{(1-r)^2}(3 - r + \frac{2\ln r}{1-r})$.
A direct insnpection of the phase dependence shows that it depends on only one combination, i.e., $\phi_2=\xi_2 +\theta_{\mu}+\chi_1 +\chi_2$.

A similar analysis to neutralino exchage contribution for the electron edm leads to the three combinations of phases, i.e.,  $\phi_1=\xi_1 +\theta_{\mu}+\chi_1 +\chi_2$,  $\phi_2=\xi_2 +\theta_{\mu}+\chi_1 +\chi_2$ and  $\phi_e=\alpha_e +\theta_{\mu}+\chi_1 +\chi_2$.
The complete set of these phases that appear in the EDMs of the electron and quarks are given in Table 1.
We use the formulae of quark and lepton edms, color edms and the purely gluonic contributions for the quarks presented in Reference ~\cite{Tarek2} and is collected in Appendix A.

Using the nonrelativistic SU(6) quark model, which connects the neutron EDM to the EDMs of its constituent quarks via the spin-flavor wavefunction, we derive the neutron electric dipole moment (EDM) from the quark EDMs.

The neutron EDM is:
\[
d_n = \frac{4}{3} d_d - \frac{1}{3} d_u
\]

Similarly, for the proton:
\[
d_p = \frac{4}{3} d_u - \frac{1}{3} d_d
\]

The contribution of the chromoelectric operator to the EDMs of quarks can be computed
using dimensional analysis~\cite{Dimensional}. The contribution to the quark EDM arising from 
$\tilde d_q^{\,C}$ is given by

\begin{equation}
	d_q^{\,C} = \frac{e}{4\pi} \, \eta^C \, \tilde d_q^{\,C}
\end{equation}

where $\eta^C \simeq 3.4$. The factor $\eta^C$ evolves the electric dipole moment
from the electroweak scale down to the hadronic scale where it can be compared
with experiment.

To obtain the contribution to the neutron and proton EDM from the quark EDM,
we use the nonrelativistic SU(6) quark model which gives

\begin{equation}
	d_n^{\,C} =
	\frac{1}{3}
	\left( 4 d_d^{\,C} - d_u^{\,C} \right).
\end{equation}

Similarly, for the proton one obtains

\begin{equation}
	d_p^{\,C} =
	\frac{1}{3}
	\left( 4 d_u^{\,C} - d_d^{\,C} \right).
\end{equation}

The contribution to $d_{n,p}$ from $d^G$ can be estimated by naive dimensional
analysis~\cite{Dimensional}, which gives

\begin{equation}
	d_{n,p}^{\,G}
	=
	\frac{e M}{4\pi}
	\, d^G \, \eta^G,
\end{equation}

where $M$ is the chiral symmetry breaking scale with numerical value
$M = 1.19 \,\mathrm{GeV}$, and $\eta^G$ is the renormalization group evolution
factor of the dimension-six operator from the electroweak scale down to the
hadronic scale.

We estimate that

\begin{equation}
	\eta^C \simeq \eta^G \sim 3.4,
\end{equation}

in agreement with the analysis of Ref.~\cite{Dimensional}.

As seen from the table, the electric dipole moments and the chromoelectric dipole moments contributions to the quarks depend on five phases which can be chosen to be $\phi_i=\xi_i +\theta_1$ and $\phi_k=\alpha_k +\theta_1$ where $\theta_1= \theta_{\mu}+\chi_1 +\chi_2$. The purely gluonic dimension six operator conribution to the quarks depends on one additional phase $\phi_t=\alpha_t +\theta_1$ and thus the nucleon EDMs depend on six independent phases. The electron EDM depends on just three indpendent phases, $\phi_i=\xi_i + \theta_1 (i=1,2)$ and $\phi_e=\alpha_e +\theta_1$.
Thus the nucleons and the electron EDMs together depend on seven independent phases.

\begin{table}[h]
	\begin{center}
		\begin{tabular}{l  c  c c}
			\hline\hline
			\multicolumn{4}{c}{\textbf{CP phases}} \\
			\hline
			exchange & u quark & d quark & electron \\
			\hline
			$\tilde{g}$ & $\phi_u=\alpha_u + \theta_1$, $\phi_3= \xi_3 +\theta_1$ \\
                                                   & &  $\phi_d=\alpha_d +\theta_1$  ,
                                                    $\phi_3=\xi_3+\theta_1$ & \\
                                 \hline
			$\chi^+$&$ \phi_d=\alpha_d +\theta_1,
                                     \phi_2= \xi_2+\theta_1$\\
                                   &&  $\phi_u=\alpha_u + \theta_1,
                                      \phi_2=\xi_2 +\theta_1$ &$\phi_2=\xi_2 + \theta_1$\\
			\hline
                   $\chi^0$&$\phi_u= \alpha_u +\theta_1$,
                                       & $\phi_d=\alpha_d + \theta_1$, & $ \phi_e= \alpha_e + \theta_1,$ \\
                                     &$ \phi_2= \xi_2+\theta_1,\phi_1=\xi_1+\theta_1$    & $\phi_2=\xi_2 +\theta_1, \phi_1=\xi_1+\theta_1$ &$\phi_2=\xi_2 + \theta_1, \phi_1=\xi_1 + \theta_1$ \\
			\hline
                     dim 6&$\phi_t=\alpha_t +\theta_1$\\
                    \hline
		\end{tabular}
		\caption{CP violating Phases in $d_q$ and $d_e$ in MSSM.}
		\label{table1}
	\end{center}
\end{table}


\section{Numerical analysis and results\label{sec4}}

In this section, we present the numerical analysis within the framework of the MSSM without assuming universal boundary conditions for the soft supersymmetry-breaking parameters, taking into account simultaneously the lightest Higgs boson mass, ($m_h \simeq 125,\mathrm{GeV}$) and the current experimental constraints on the electric dipole moments of the electron, neutron and proton.

\begin{figure}[H]
	\begin{center}
		{\rotatebox{0}{\resizebox*{7.5cm}{!}{\includegraphics{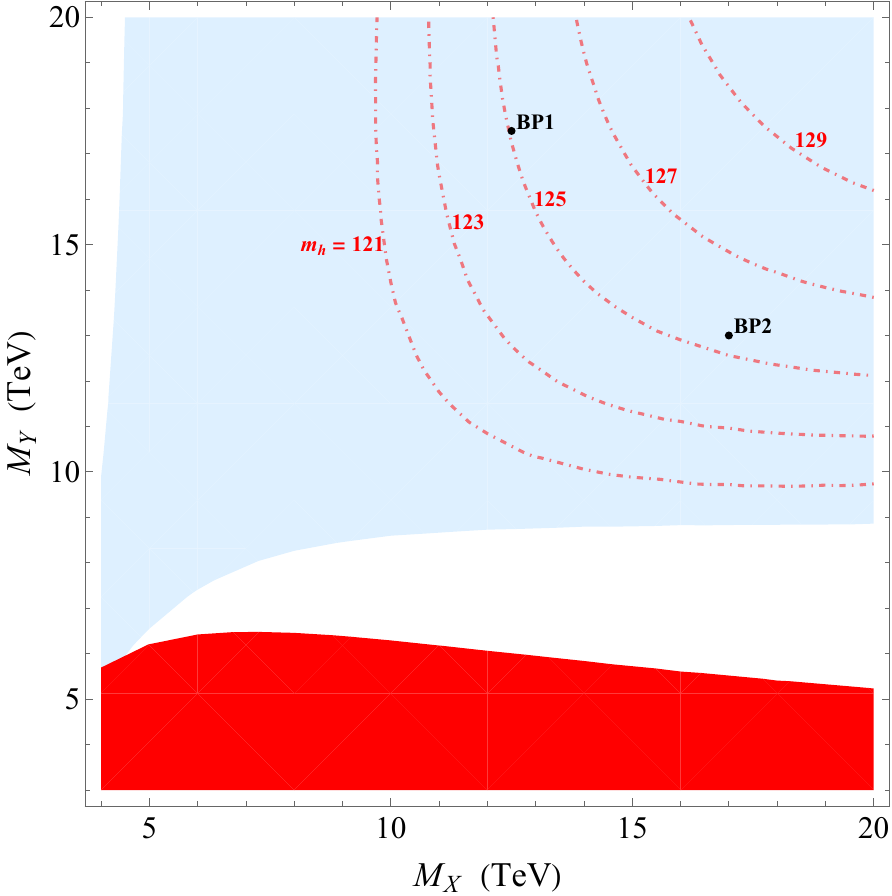}}\hglue5mm}}
		{\rotatebox{0}{\resizebox*{7.5cm}{!}{\includegraphics{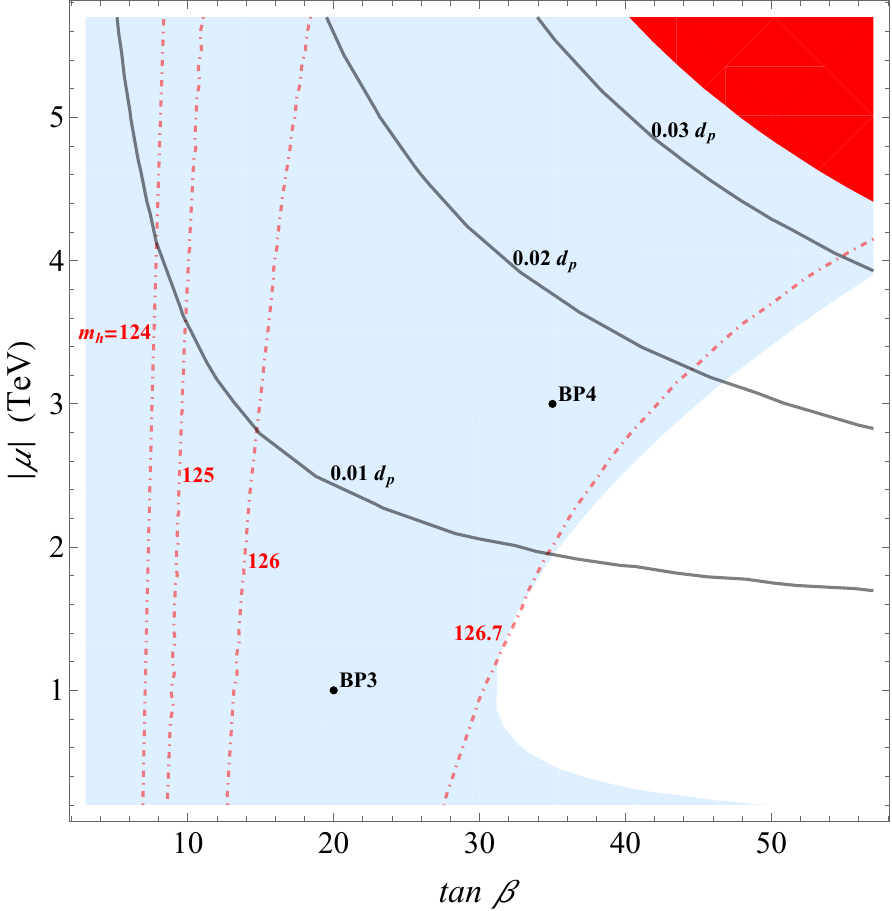}}\hglue5mm}}\\
		\caption{Left panel:
			Allowed regions in the $(M_X,\,M_Y)$ plane consistent with the current experimental upper bound on the electric dipole moments. ($\tan\beta = 10$, $|\mu|=500$, $ M_{\tilde Q} = M_Y $,, $ M_{\tilde U} = M_X $,  $M_{\tilde D}=M_X$,  $M_{\tilde R_\ell}=2M_X$, $M_{\tilde L_\ell}=6M_Y$, $|A_u|=|A_d|=|A_e|=2M_X$, and $|A_t|=2(M_X+M_Y)$). 
			Right panel: Allowed regions in the $(\tan\beta,\,|\mu|)$ plane consistent with the current experimental upper bound on the electric dipole moments. ($M_{\tilde U}=M_{\tilde D}=17$, $M_{\tilde Q}=13$, $M_{\tilde R_\ell}=17$, $M_{\tilde L_\ell}=83$, $|A_u|=|A_d|=|A_e|=34$, and $|A_t|=60$). The blue regions represent the parameter space allowed by the current upper limit on the electron EDM, while the red regions are excluded by the current upper limit on the neutron EDM. The red dot-dashed contours in both panels denote the lightest Higgs boson mass. The black contours in the right panel correspond to 1 \%, 2 \%, and 3 \% of the current experimental upper limit on the proton electric dipole moment. Other parameters have the values $|M_1| = 0.2$, $|M_2| = 2$, $m_{\tilde{g}} = 2.7 $, $m_A=0.5$, $\phi_1 = 1.7$, $\phi_2 = 3.2$, $\phi_3 = 1.7$, $\phi_u = 0.7$, $\phi_d = 0.7$, $\phi_t = 1.7$, and $\phi_e = 1.2$. All masses are in TeV and all phases in rad.}
		\label{figure01}
	\end{center}
\end{figure}

Figure~\ref{figure01} shows the phenomenologically viable regions of the MSSM parameter space after imposing the current experimental constraints on the electric dipole moments. The left panel displays the allowed region in the $(M_X,M_Y)$ plane, where $M_X \equiv M_{\tilde U}=M_{\tilde D}$ and $M_Y \equiv M_{\tilde Q}$ denote the common right-handed and left-handed squark soft masses, respectively. In this scan, the remaining scalar masses and trilinear couplings are chosen according to $M_{\tilde R_\ell}=2M_X$, $M_{\tilde L_\ell}=6M_Y$, $|A_u|=|A_d|=|A_e|=2M_X$, and $|A_t|=2(M_X+M_Y)$, while $\tan\beta=10$ and $|\mu|=500~\mathrm{GeV}$ are fixed.
 The right panel shows the allowed region in the $(\tan\beta,|\mu|)$ plane for the parameter set specified in the caption.
The blue regions satisfy the present upper limit on the electron EDM, whereas the red regions are excluded by the neutron EDM constraint. The contours of the lightest Higgs boson mass are shown in both panels, while the right panel also includes contours corresponding to the projected proton EDM sensitivity.
The combined Higgs boson mass and EDM constraints significantly restrict the MSSM parameter space, leaving only regions that simultaneously reproduce the observed Higgs boson mass and satisfy the current experimental EDM bounds.

\begin{table}[H]
	\centering
	\begin{tabular}{l c c c c}
		\hline\hline
		& BP1 & BP2 & BP3 & BP4 \\
		\hline	
		$m_h$ (GeV)	& $125.07$ & $125.45$ & $126.48$ & $126.65$ \\
		\hline\hline
		$|d_e|$ & $9.37 \times 10^{-31}$ & $1.88 \times 10^{-30}$ & $2.65 \times 10^{-30}$ & $ 3.19 \times 10^{-30}$ \\
		\hline\hline
		$d_n^{E}$ & $1.39 \times 10^{-27}$ & $2.23 \times 10^{-27}$ & $3.48 \times 10^{-27}$ & $2.66 \times 10^{-27}$ \\
		$d_n^{C}$  & $1.23 \times 10^{-27}$ & $1.82 \times 10^{-27}$ & $4.99 \times 10^{-28}$ & $-6.98 \times 10^{-27}$ \\
		$d_n^{G}$  & $1.96 \times 10^{-27}$ & $1.95 \times 10^{-27}$ & $1.95 \times 10^{-27}$ & $1.95 \times 10^{-27}$ \\
		\hline
		Total $|d_n|$ & $4.58 \times 10^{-27}$ & $6.00 \times 10^{-27}$ & $5.93 \times 10^{-27}$ & $ 2.37 \times 10^{-27}$ \\
		\hline\hline
		$d_p^{E}$ & $-9.62 \times 10^{-28}$ & $-1.38 \times 10^{-27}$ & $-1.69 \times 10^{-27}$ & $-1.48 \times 10^{-27}$ \\
		$d_p^{C}$  & $5.06 \times 10^{-28}$ & $6.42 \times 10^{-28}$ & $9.71 \times 10^{-28}$ & $2.84 \times 10^{-27}$ \\
		$d_p^{G}$  & $1.96 \times 10^{-27}$ & $1.95 \times 10^{-27}$ & $1.95 \times 10^{-27}$ & $1.95 \times 10^{-27}$ \\
		\hline
		Total $|d_p|$ & $1.50 \times 10^{-27}$ & $1.21 \times 10^{-27}$ & $1.23 \times 10^{-27}$ & $ 3.31 \times 10^{-27}$ \\
		\hline\hline
	\end{tabular}
	\caption{Electron, Neutron, and Proton EDMs for the four benchmark points considered in figure~\ref{figure01}. The inputs are listed in table~\ref{BPinputs}. All values of EDM are given in units of $e\mathrm{cm}$.}
	\label{table01_edm}
\end{table}

Four representative benchmark points, listed in tables~\ref{table01_edm}, are chosen from the allowed parameter space of figure~\ref{figure01}. The benchmark points satisfy the Higgs boson mass constraint while remaining consistent with the current experimental limits on the electron, neutron, and proton EDMs. In table~\ref{table01_edm}, we present the individual electric, chromoelectric and purely gluonic contributions to the neutron and proton EDMs, together with their total values. One finds that the purely gluonic contribution is the dominant component for all four benchmark points, whereas the electric and chromoelectric contributions exhibit a stronger dependence on the SUSY parameters. The total neutron and proton EDMs arise from the sum of these three contributions, and their numerical values reflect the constructive or destructive interference among the different CP-violating operators. The benchmark points therefore demonstrate that sizable CP-violating phases can remain compatible with the current experimental constraints through the interplay of the various EDM contributions while simultaneously reproducing the observed Higgs boson mass.

The MSSM sector inputs of the four benchmark points in table~\ref{table01_edm} are shown in table~\ref{BPinputs}.
\begin{table}[H]
	\begin{center}
		\begin{tabular}{l c c c c c c c c c c c }
			\hline\hline
			& $\tan\beta$ & $|\mu|$ & $M_{\tilde Q}$ & $M_{\tilde U}$ & $M_{\tilde D}$ & $M_{\tilde L_\ell}$ & $M_{\tilde R_\ell}$ & $|A_u|$ & $|A_d|$ & $|A_t|$ & $|A_e|$\\
			\hline
			BP1 & $10$ & $0.5$ & $17.5$ & $12.5$ & $12.5$ & $105$ & $25$ & $25$ & $25$ & $60$ & $25$ \\
			BP2 & $10$ & $0.5$ & $13$ & $17$ & $17$ & $78$ & $34$ & $34$ & $34$ & $60$ & $34$ \\
			BP3 & $20$ & $1$ & $13$ & $17$ & $17$ & $83$ & $17$ & $34$ & $34$ & $60$ & $34$ \\
			BP4 & $35$ & $3$ & $13$ & $17$ & $17$ & $83$ & $17$ & $34$ & $34$ & $60$ & $34$ \\
			\hline
		\end{tabular}
		\caption{Input parameters of the four benchmark points of table~\ref{table01_edm}. The remaining MSSM input parameters are common to all benchmark points and are fixed as $|M_1| = 0.2$, $|M_2| = 2$, $m_{\tilde{g}} = 2.7 $, $m_A=0.5$, $\phi_1 = 1.7$, $\phi_2 = 3.2$, $\phi_3 = 1.7$, $\phi_u = 0.7$, $\phi_d = 0.7$, $\phi_t = 1.7$, and $\phi_e = 1.2$. All masses are in TeV and all phases in rad.}
		\label{BPinputs}
	\end{center}
\end{table}

The benchmark points listed in Table~\ref{BPinputs} lead to the corresponding physical supersymmetric spectrum after diagonalization of the relevant mass matrices. The resulting masses of the lightest neutralino, lightest chargino, gluino, lightest stop, lightest up-type squark, lightest down-type squark, and lightest selectron are presented in Table~\ref{physical_masses}. The spectrum is characterized by a relatively light neutralino, while the colored
scalar sector remains at the multi-TeV scale. In particular, the squark and stop masses are around $13$ TeV, whereas the gluino mass is $2.7$ TeV.
These colored-particle masses are well above the exclusion ranges obtained in recent ATLAS and CMS searches for gluinos, squarks, and top squarks in the corresponding simplified scenarios \cite{ATLAS:2025SquarkGluinoTau,ATLAS:2026StopDilepton,CMS:2025TopSquarkJets}.
The lightest neutralino has a mass of $m_{\tilde{\chi}_1^0}=0.20$ TeV and can therefore serve as a viable dark-matter candidate in the MSSM. The relic abundance, however, depends on its composition and annihilation mechanisms and requires a separate analysis ~\cite{Planck:2018Cosmo}.

\begin{table}[H]
	\begin{center}
		\begin{tabular}{lccccccc}
			\hline\hline
			&
			$m_{\tilde{\chi}_1^0}$ &
			$m_{\tilde{\chi}_1^\pm}$ &
			$m_{\tilde g}$ &
			$m_{\tilde t_1}$ &
			$m_{\tilde u_1}$ &
			$m_{\tilde d_1}$ &
			$m_{\tilde e_1}$ \\
			\hline
			{BP1} &$0.20$ &$0.50$ &$2.70$ &$12.47$ &$12.50$ &$12.50$ &$25.00$ \\
			{BP2} &$0.20$ &$0.50$ &$2.70$ &$12.97$ &$13.00$ &$13.00$ &$34.00$ \\
			{BP3} &$0.20$ &$1.00$ &$2.70$ &$12.97$ &$13.00$ &$13.00$ &$17.00$ \\
			{BP4} &$0.20$ &$2.00$ &$2.70$ &$12.97$ &$13.00$ &$13.00$ &$17.00$ \\
			\hline
		\end{tabular}
		
		\caption{Physical masses of the lightest neutralino ($\tilde{\chi}_1^0$), lightest chargino ($\tilde{\chi}_1^\pm$), gluino ($\tilde g$), lightest stop ($\tilde t_1$), lightest up squark ($\tilde u_1$), lightest down squark ($\tilde d_1$), and lightest selectron ($\tilde e_1$) for the four benchmark points listed in Table~\ref{BPinputs}. All masses are given in TeV.}
		\label{physical_masses}
	\end{center}
\end{table}

\begin{figure}[H]
	\begin{center}
		{\rotatebox{0}{\resizebox*{7.3cm}{!}{\includegraphics{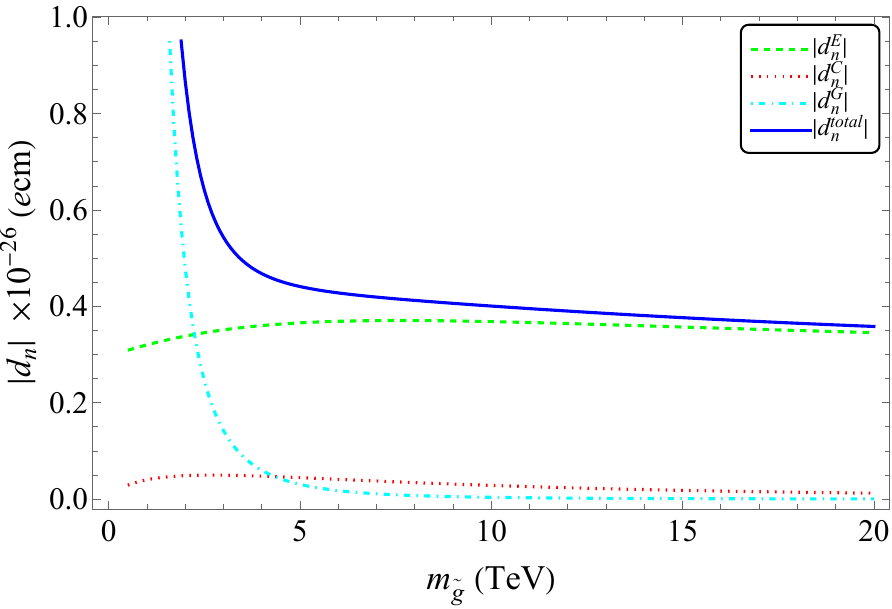}}\hglue5mm}}
		{\rotatebox{0}{\resizebox*{7.3cm}{!}{\includegraphics{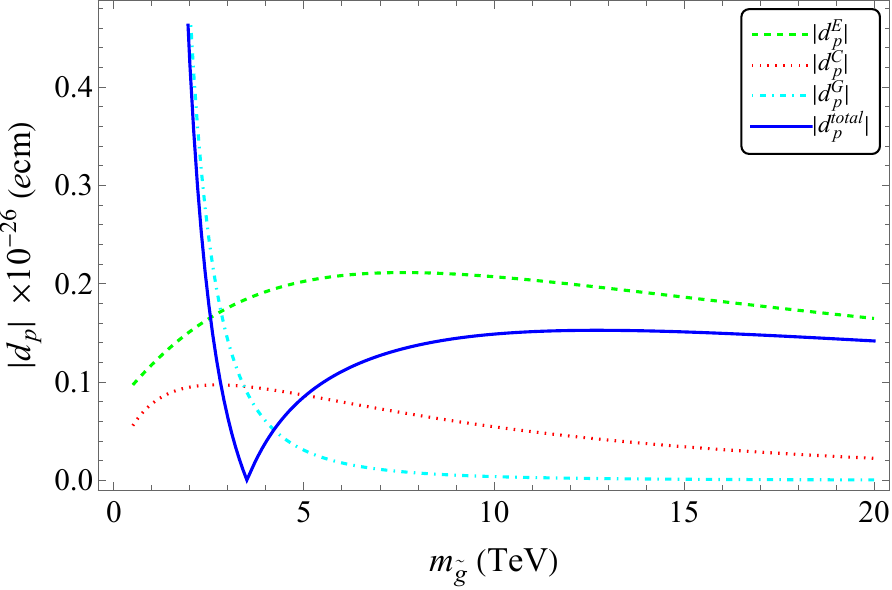}}\hglue5mm}}\\
		\caption{Left panel: Variation of the neutron EDM, $|d_n|$ versus the gluino mass, $ m_{\tilde{g}}$. Right panel: Variation of the proton EDM, $|d_p|$ versus the gluino mass, $ m_{\tilde{g}} $. The parameter values correspond to Benchmark Point 3 of table~\ref{BPinputs}.}
		\label{dn-vs-mg}
	\end{center}
\end{figure}

The nearly unchanged values of the purely gluonic contribution in Table~\ref{table01_edm} are a consequence of the similar gluino and stop-sector parameters adopted for the selected benchmark points. To demonstrate the sensitivity of this contribution, Fig.~\ref{dn-vs-mg} displays the neutron and proton EDMs as functions of the gluino mass $m_{\tilde{g}}$ for Benchmark Point~3. As expected, increasing the gluino mass suppresses the purely gluonic contribution, resulting in smaller neutron and proton EDMs. 


\begin{figure}[H]
	\begin{center}
		{\rotatebox{0}{\resizebox*{7.3cm}{!}{\includegraphics{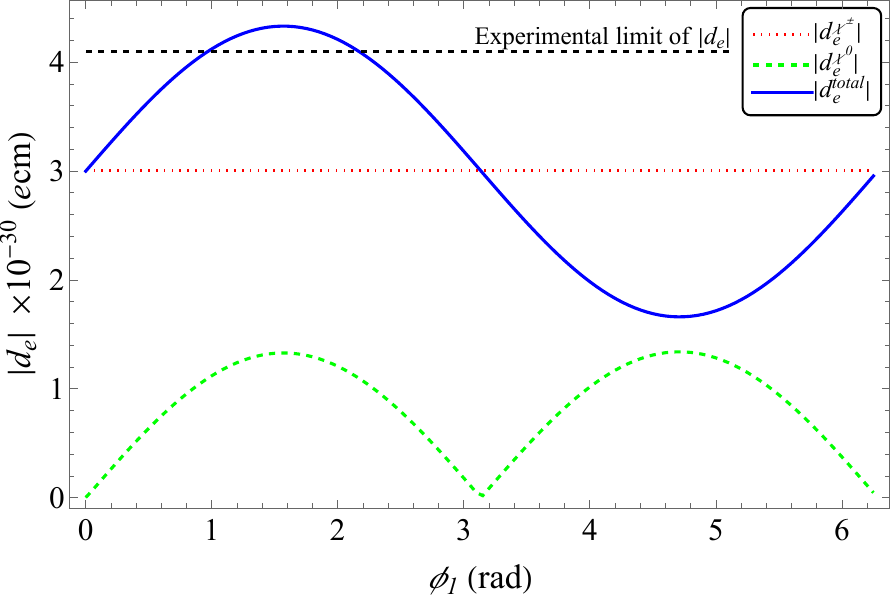}}\hglue5mm}}
		{\rotatebox{0}{\resizebox*{7.3cm}{!}{\includegraphics{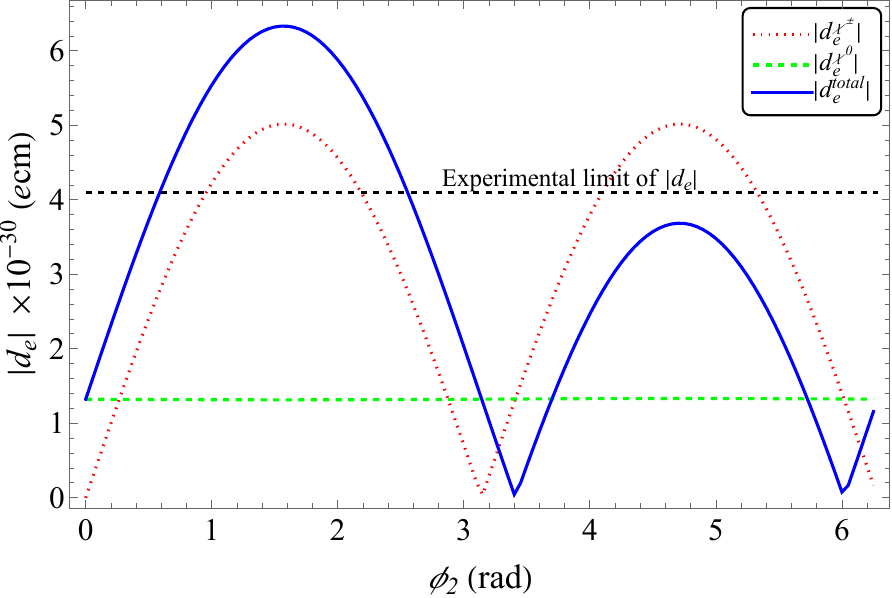}}\hglue5mm}}\\
		\caption{Left panel: Variation of the electron EDM, $|d_e|$ (solid curve), the chagino contribution $|d_e^{\chi^\pm}|$ (dotted curve), and the neutralino contribution $|d_e^{\chi^0}|$ (dashed curve), versus $ \phi_1 $, for $ \phi_2 = 2.5 $. Right panel: Variation of the electron EDM, $|d_e|$ (solid curve), the chagino contribution $|d_e^{\chi^\pm}|$ (dotted curve), and the neutralino contribution $|d_e^{\chi^0}|$ (dashed curve), versus $ \phi_2 $, for $ \phi_1 = 1.7 $. Other parameters have the values: $ \tan\beta = 10 $,	$|\mu| = 0.5$, $|M_1| = 0.2$, $|M_2| = 2$, $M_{\tilde L_\ell}=200$, $M_{\tilde R_\ell}=13$, $|A_e| = 25 $, and $\phi_e = 1.2$. All masses are in TeV and all phases in rad.}
		\label{figure02}
	\end{center}
\end{figure}

The dependence of the electron EDM on the electroweak gaugino phases is illustrated in Fig.~\ref{figure02}. The left and right panels show the variation of the total electron EDM together with its chargino and neutralino contributions as functions of $\phi_1$ and $\phi_2$, respectively. Both contributions exhibit a pronounced dependence on the CP-violating phases, and their constructive or destructive interference determines the magnitude of the total electron EDM.

\begin{figure}[H]
	\begin{center}
		{\rotatebox{0}{\resizebox*{7.3cm}{!}{\includegraphics{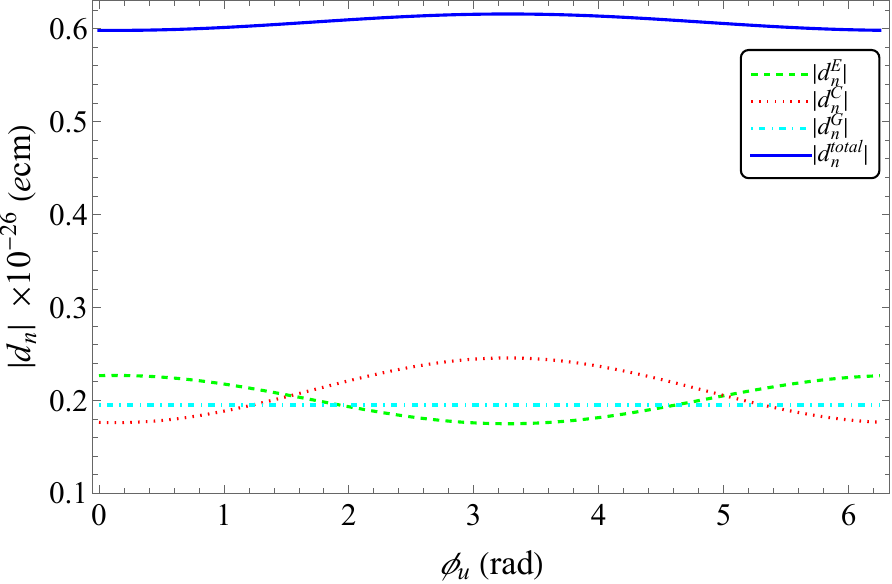}}\hglue5mm}}
		{\rotatebox{0}{\resizebox*{7.3cm}{!}{\includegraphics{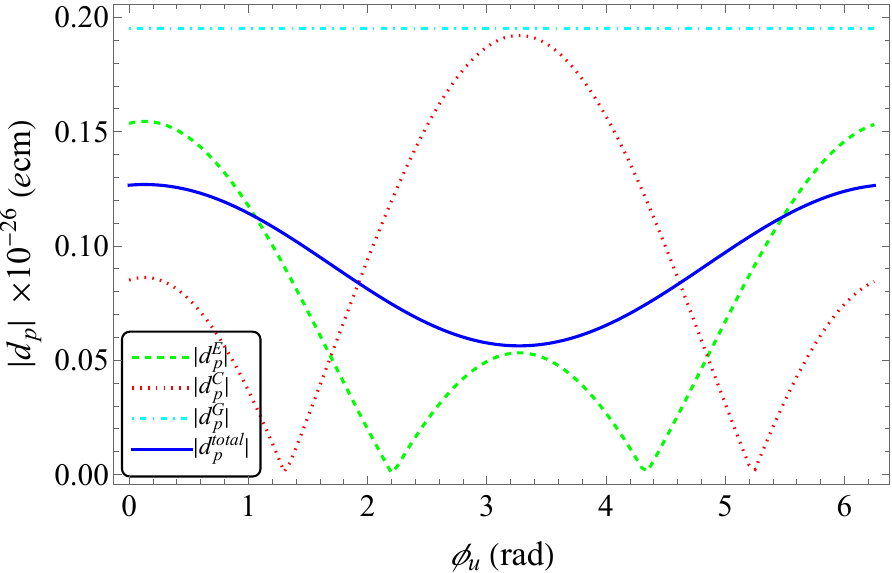}}\hglue5mm}}\\
		\caption{Left panel: Variation of the neutron EDM, $|d_n|$ versus $ \phi_u $. Right panel: Variation of the proton EDM, $|d_p|$ versus $ \phi_u $. The parameter values correspond to Benchmark Point 2 (BP2) of table~\ref{BPinputs}.}
		\label{figure03}
	\end{center}
\end{figure}

\begin{figure}[H]
	\begin{center}
		{\rotatebox{0}{\resizebox*{7.3cm}{!}{\includegraphics{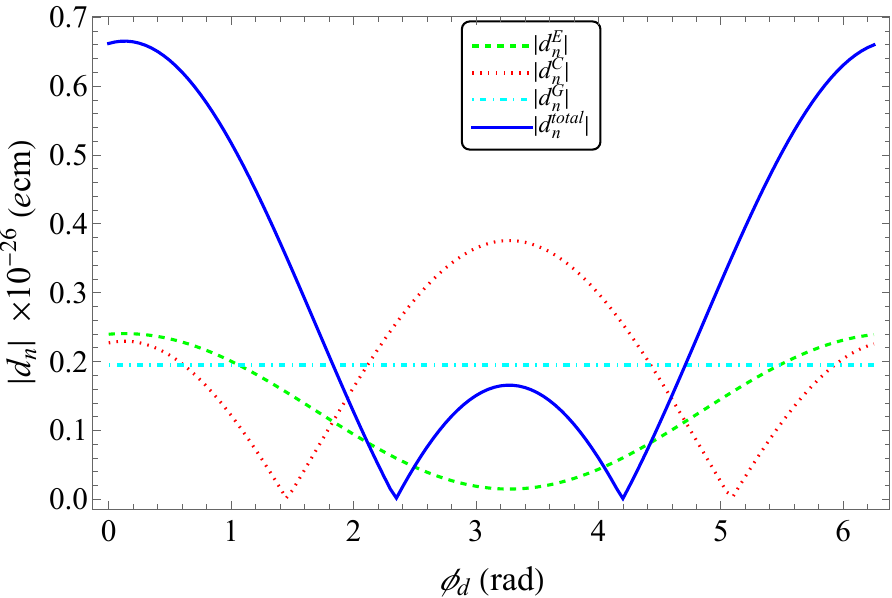}}\hglue5mm}}
		{\rotatebox{0}{\resizebox*{7.3cm}{!}{\includegraphics{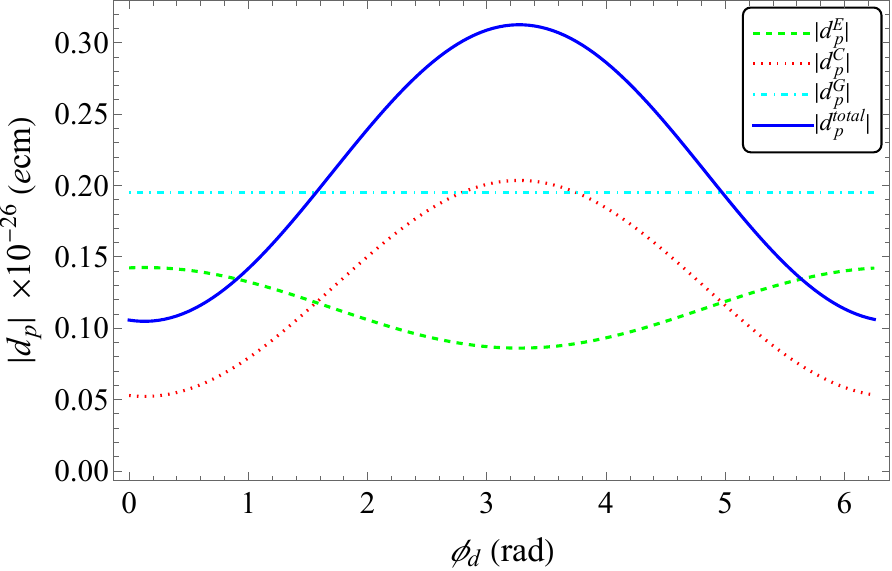}}\hglue5mm}}\\
		\caption{Left panel: Variation of the neutron EDM, $|d_n|$ versus $ \phi_d $. Right panel: Variation of the proton EDM, $|d_p|$ versus $ \phi_d $. The parameter values correspond to Benchmark Point 2 of table~\ref{BPinputs}.}
		\label{figure04}
	\end{center}
\end{figure}

Figures~\ref{figure03} and \ref{figure04} show the dependence of the neutron and proton EDMs on the phase $\alpha_u$ and $\alpha_d$  respectively. Both observables exhibit a strong sensitivity to these phases, although their numerical behavior differs due to the different combinations of the electric, chromoelectric and purely gluonic contributions entering each EDM.

\begin{figure}[H]
	\begin{center}
		{\rotatebox{0}{\resizebox*{7.3cm}{!}{\includegraphics{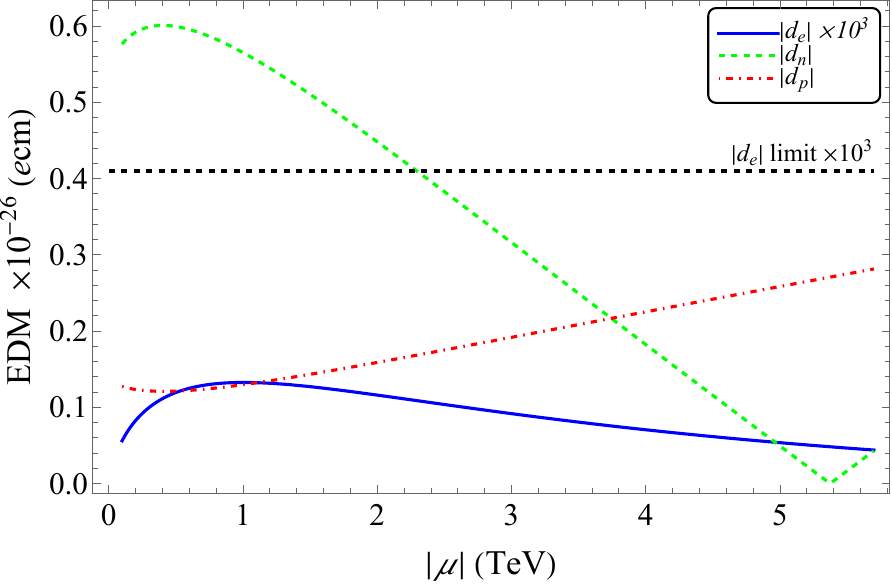}}\hglue5mm}}
		{\rotatebox{0}{\resizebox*{7.3cm}{!}{\includegraphics{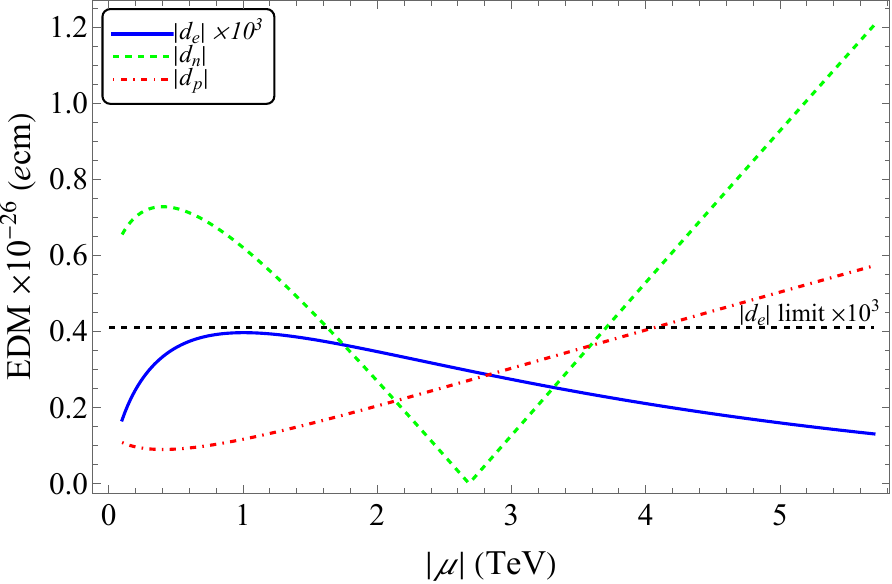}}\hglue5mm}}\\
		\caption{Variation of the electron, neutron, and proton electric dipole moments versus the Higgsino mass parameter $|\mu|$, for $\tan\beta = 10$ (left panel) and $\tan\beta = 30$ (right panel). The remaining MSSM input parameters are identical to those used in the right panel of Fig.~\ref{figure01}. The electron EDM has been multiplied by a factor of $10^3$ for clarity.}
		\label{figure05}
	\end{center}
\end{figure}

Figure~\ref{figure05} illustrates the dependence of the electron, neutron, and proton EDMs on the Higgsino mass parameter $|\mu|$ for two representative values of $\tan\beta$. The three observables exhibit distinct dependences on $|\mu|$, reflecting the different combinations of supersymmetric contributions entering each EDM. While the electron EDM varies smoothly over the considered range of $|\mu|$, the neutron and proton EDMs display a more pronounced dependence due to the interplay among the electric, chromoelectric, and purely gluonic contributions.
For $\tan\beta=10,(30)$, the neutron EDM exhibits a pronounced minimum around $|\mu|\simeq5,(3)~\mathrm{TeV}$, respectively, indicating a strong cancellation among the different contributions.
No comparable cancellation is observed for the proton EDM, which remains sizable throughout the scanned region. Furthermore, these results highlight the important role of the Higgsino mass parameter in shaping the EDM predictions and complement the $(\tan\beta,|\mu|)$ analysis presented in Fig.~\ref{figure01}.

Overall, the numerical analysis demonstrates that the simultaneous Higgs-mass and EDM constraints significantly restrict the MSSM parameter space while still allowing viable benchmark scenarios with sizeable CP-violating phases. Future improvements in proton EDM measurements will further enhance the sensitivity to these scenarios and provide an important complementary probe of CP violation in supersymmetry.


\newpage





\section{Conclusion\label{sec5}}
In this paper we have presented the analysis of the Higgs mass calculation by the tree level and by the loop correction of the scalar potential of the MSSM. The discovery of the Higgs boson mass at 125 GeV raises the SUSY Breaking parameters into the TeV scale. However, some of these parameters could generally be complex and we have collected their phases into seven specific sets that enter the analysis of the electric dipole moments.
 For the case of minimal supergravity,  the seven phases evolve from only two phases at the GUT scale, $\phi_1=\alpha_0 +\theta_1 $ and $\phi_2=\xi_{1/2}+ \theta_1$ where $\xi_{1/2}$ and $\alpha_0$ are the gaugino phase and trilinear coupling phase at the GUT scale.
 We have also presented the analysis of the EDMs of the electron, the neutron and the proton within the same framework of MSSM where the seven phases are not constrained. For the electron, we have considered the chargino and neutralino contributrions. For the neutron and proton we considered also the contributions from the gluino, the chromoelectric and from the purely gluonic operators. One finds that by relaxing the CP violating phases to be O(1), as there is no theoretical reason that would fine tune them, we have found that the combined Higgs boson mass and EDM constraints significantly restrict the MSSM parameter space, leaving only regions that simultaneously reproduce the observed Higgs boson mass and satisfy the current experimental EDM bounds. In these allowed regions, although the discovery of the SUSY partners would be very difficult in the current  colliders, the low energy probes of the EDMs of the electron, proton and neutron can open windows for these heavy partners detection as they are running in the loops that produce the different components of the electric dipole moments.
\\
\noindent

\section{Appendix A: EDM components of quarks and leptons\label{sec6}}


The Electric Dipole Moment (EDM) of a spin-1/2 particle is defined by the effective Lagrangian:

\begin{equation}
	L_I = -\frac{i}{2} d_f \bar{\psi} \sigma^{\mu\nu} \gamma_5 \psi F_{\mu\nu}
	\label{eq:edm_lagrangian}
\end{equation}

where $d_f$ is the EDM of the fermion $f$, $\psi$ is the fermion field, $\sigma^{\mu\nu} = \frac{i}{2}[\gamma^{\mu}, \gamma^{\nu}]$, $\gamma_5$ is the chiral gamma matrix, and $F_{\mu\nu}$ is the electromagnetic field strength tensor. The contributions to the fermion EDMs arise from various supersymmetric particle exchanges, including gluinos, charginos, and neutralinos.

\subsubsection*{A.1. Gluino Contribution to Quark EDM}

The gluino exchange generates a contribution to the electric dipole moment (EDM) of a quark, which is given by

\begin{equation}
	d^E_{q-\mathrm{gluino}} / e =
	-\frac{2\alpha_s}{3\pi}
	\frac{m_{\tilde g}}
	{M_{\tilde q_1}^2-M_{\tilde q_2}^2}
	\,
	\mathrm{Im}\!\left(\Gamma_q^{11}\right)
	\left[
	\frac{1}{M_{\tilde q_1}^2}
	B\!\left(\frac{m_{\tilde g}^2}{M_{\tilde q_1}^2}\right)
	-
	\frac{1}{M_{\tilde q_2}^2}
	B\!\left(\frac{m_{\tilde g}^2}{M_{\tilde q_2}^2}\right)
	\right],
	\label{eq:gluino_edm}
\end{equation}

where $\alpha_s$ denotes the strong coupling constant, $m_{\tilde g}$ is the gluino mass, and $M_{\tilde q_1}$ and $M_{\tilde q_2}$ are the physical masses of the two squark mass eigenstates. The loop function $B(r)$ is defined as

\begin{equation}
	B(r)=
	\frac{1}{2(r-1)^2}
	\left(
	1+r+\frac{2r\ln r}{1-r}
	\right),
	\label{eq:Bloop}
\end{equation}

The CP-violating contribution enters through the imaginary part of the effective quark--gluino--squark coupling,

\begin{equation}
	\mathrm{Im}\!\left(\Gamma_q^{11}\right)
	=
	\frac{m_q}
	{M_{\tilde q_1}^2-M_{\tilde q_2}^2}
	\left[
	|A_q|
	\sin(\alpha_q-\xi_3)
	+
	|\mu||R_q|
	\sin(\xi_3+\theta_{\mu}+\chi_1 +\chi_2)
	\right],
	\label{eq:gamma_11q}
\end{equation}

where $m_q$ is the quark mass, $A_q$ is the trilinear soft SUSY-breaking coupling, $\mu$ is the Higgsino mass parameter. The factor $R_q$ is defined by $R_q=v_1/v_2^{*}\,(v_2/v_1^{*})$ for $q=u\,(d)$, corresponding to $|R_u|=\cot\beta$ and $|R_d|=\tan\beta$.

\subsubsection*{A.2. Chargino Contribution to Quark and Lepton EDMs}

The chargino exchange generates a contribution to the electric dipole moment (EDM) of the up quark, which is given by

\begin{equation}
	d^E_{u-\mathrm{chargino}} / e =
	-\frac{\alpha_{EM}}
	{4\pi \sin^2\theta_W}
	\sum_{k=1}^{2}
	\sum_{i=1}^{2}
	\frac{m_{\tilde{\chi}_i^+}}
	{M_{\tilde d_k}^{\,2}}
	\,\mathrm{Im}\!\left(\Gamma_{uik}\right)
	\left[
	Q_{\tilde d}
	B\!\left(
	\frac{m_{\tilde{\chi}_i^+}^{2}}
	{M_{\tilde d_k}^{\,2}}
	\right)
	+
	(Q_u-Q_{\tilde d})
	A\!\left(
	\frac{m_{\tilde{\chi}_i^+}^{2}}
	{M_{\tilde d_k}^{\,2}}
	\right)
	\right],
	\label{eq:chargino_quark_edm}
\end{equation}

where $\alpha_{EM}$ is the electromagnetic coupling constant, $\theta_W$ is the weak mixing angle, $m_{\tilde{\chi}_i^+}$ is the mass of the $i$th chargino, $M_{\tilde d_k}$ is the mass of the $k$th down-type squark, and $Q_u$ and $Q_{\tilde d}$ denote the electric charges of the up quark and down-type squark, respectively. The loop function $A(r)$ is defined as

\begin{equation}
	A(r)=
	\frac{1}{2(1-r)^2}
	\left(
	3-r+\frac{2\ln r}{1-r}
	\right),
	\label{eq:Aloop}
\end{equation}

while the loop function $B(r)$ is given in Eq.~(\ref{eq:Bloop}). The effective chargino--quark--squark coupling is

\begin{equation}
	\Gamma_{uik}
	=
	\kappa_u
	V_{i2}^{*}
	D_{d\,1k}
	\left(
	U_{i1}^{*}D_{d\,1k}^{*}
	-
	\kappa_d
	U_{i2}^{*}
	D_{d\,2k}^{*}
	\right),
	\label{eq:Gamma_uik}
\end{equation}

where

\begin{equation}
	\kappa_u=e^{-i\chi_1}
	\frac{m_u }
	{\sqrt{2}m_W\sin\beta},
	\qquad
	\kappa_{d,e}=e^{-i\chi_2}
	\frac{m_{d,e}}
	{\sqrt{2}m_W\cos\beta}.
	\label{eq:kappa}
\end{equation}

For charged leptons, the chargino contribution to the EDM is

\begin{equation}
	d^E_{e-\mathrm{chargino}} / e =
	-\frac{\alpha_{EM}}
	{4\pi\sin^2\theta_W}
	\sum_{i=1}^{2}
	\frac{m_{\tilde{\chi}_i^+}}
	{M_{\tilde{\nu}_e}^{\,2}}
	\,
	\mathrm{Im}\!\left(
	\kappa_e
	U_{i2}^{*}
	V_{i1}
	\right)
	A\!\left(
	\frac{m_{\tilde{\chi}_i^+}^{2}}
	{M_{\tilde{\nu}_e}^{\,2}}
	\right),
	\label{eq:chargino_lepton_edm}
\end{equation}

where $M_{\tilde{\nu}_e}$ is the electron sneutrino mass. Further details regarding the couplings $\Gamma_{uik}$, the chargino and squark mixing matrices, and their dependence on the CP-violating phases can be found in Ref.~\cite{Tarek2}.

\subsubsection*{A.3. Neutralino Contribution to Fermion EDM}

The neutralino exchange generates a contribution to the electric dipole moment (EDM) of a fermion, which is given by

\begin{equation}
	d^E_{f-\mathrm{neutralino}} / e =
	-\frac{\alpha_{EM}}
	{4\pi\sin^2\theta_W}
	\sum_{k=1}^{2}
	\sum_{i=1}^{4}
	\frac{m_{\tilde{\chi}_i^0}}
	{M_{\tilde f_k}^{\,2}}
	\,
	\mathrm{Im}\!\left(\eta_{fik}\right)
	Q_{\tilde f}
	B\!\left(
	\frac{m_{\tilde{\chi}_i^0}^{2}}
	{M_{\tilde f_k}^{\,2}}
	\right),
	\label{eq:neutralino_edm}
\end{equation}

where $\alpha_{EM}$ is the electromagnetic coupling constant, $m_{\tilde{\chi}_i^0}$ is the mass of the $i$th neutralino, $M_{\tilde f_k}$ is the mass of the $k$th sfermion mass eigenstate, and $Q_{\tilde f}$ is the electric charge of the sfermion. The loop function $B(r)$ is defined in Eq.~(\ref{eq:Bloop}). The effective neutralino--fermion--sfermion coupling $\eta_{fik}$ depends on the neutralino and sfermion mixing matrices and the CP-violating phases. Its explicit form can be found in Ref.~\cite{Tarek2}.

\subsection*{B. Chromoelectric Dipole Moments}

The chromoelectric dipole moment (CEDM) of a quark is described by the effective dimension-five operator

\begin{equation}
	\mathcal{L}_{\mathrm{CEDM}}
	=
	-\frac{i}{2}
	\tilde d_q^C
	\,
	\bar q
	\sigma_{\mu\nu}\gamma_5
	T^a
	q
	G^{\mu\nu a},
	\label{eq:cedm_lagrangian}
\end{equation}

where $T^a$ are the generators of the SU(3)$_C$ gauge group and $G_{\mu\nu}^a$ is the gluon field-strength tensor. In the MSSM, the CEDM receives contributions from gluino, chargino, and neutralino exchange diagrams.

The gluino contribution is given by

\begin{equation}
	\tilde d^C_{q-\mathrm{gluino}}
	=
	\frac{g_s\alpha_s}{4\pi}
	\sum_{k=1}^{2}
	\frac{m_{\tilde g}}
	{M_{\tilde q_k}^{\,2}}
	\,
	\mathrm{Im}\!\left(\Gamma_q^{1k}\right)
	C\!\left(
	\frac{m_{\tilde g}^{\,2}}
	{M_{\tilde q_k}^{\,2}}
	\right),
	\label{eq:gluino_cedm}
\end{equation}

The chargino contribution is

\begin{equation}
	\tilde d^C_{q-\mathrm{chargino}}
	=
	-\frac{g^2g_s}{16\pi^2}
	\sum_{k=1}^{2}
	\sum_{i=1}^{2}
	\frac{m_{\tilde\chi_i^+}}
	{M_{\tilde q_k}^{\,2}}
	\,
	\mathrm{Im}\!\left(\Gamma_{qik}\right)
	B\!\left(
	\frac{m_{\tilde\chi_i^+}^{\,2}}
	{M_{\tilde q_k}^{\,2}}
	\right),
	\label{eq:chargino_cedm}
\end{equation}

while the neutralino contribution is

\begin{equation}
	\tilde d^C_{q-\mathrm{neutralino}}
	=
	\frac{g_sg^2}{16\pi^2}
	\sum_{k=1}^{2}
	\sum_{i=1}^{4}
	\frac{m_{\tilde\chi_i^0}}
	{M_{\tilde q_k}^{\,2}}
	\,
	\mathrm{Im}\!\left(\eta_{qik}\right)
	B\!\left(
	\frac{m_{\tilde\chi_i^0}^{\,2}}
	{M_{\tilde q_k}^{\,2}}
	\right),
	\label{eq:neutralino_cedm}
\end{equation}

where the loop function $B(r)$ is defined in Eq.~(\ref{eq:Bloop}), while

\begin{equation}
	C(r)
	=
	\frac{1}{6(r-1)^2}
	\left(
	10r-26
	+\frac{2r\ln r}{1-r}
	-\frac{18\ln r}{1-r}
	\right).
	\label{eq:Cloop}
\end{equation}

The CP-violating phases are entirely contained in the quantities
$\mathrm{Im}(\Gamma_q^{1k})$,
$\mathrm{Im}(\Gamma_{qik})$,
and
$\mathrm{Im}(\eta_{qik})$,
which are the same factors appearing in the corresponding gluino, chargino, and neutralino contributions to the electric dipole moment discussed in Sec.~A.

\subsection*{C. Purely Gluonic Dimension-Six Operator}

The gluonic dipole moment $d^{G}$ is defined through the effective dimension-six operator

\begin{equation}
	\mathcal{L}_{\mathrm{G}}
	=
	-\frac{1}{6}
	d^{G}
	f_{\alpha\beta\gamma}
	G^{\alpha}_{\mu\rho}
	G^{\beta\,\rho}_{\ \ \nu}
	G^{\gamma}_{\lambda\sigma}
	\epsilon^{\mu\nu\lambda\sigma},
	\label{eq:weinberg_operator}
\end{equation}

where $G_{\mu\nu}^{\alpha}$ is the gluon field-strength tensor, $f_{\alpha\beta\gamma}$ are the SU(3) structure constants, and $\epsilon^{\mu\nu\lambda\sigma}$ is the totally antisymmetric Levi--Civita tensor with $\epsilon^{0123}=+1$.

The dominant contribution to $d^{G}$ arises from the top quark--stop loop with gluino exchange and is given by

\begin{equation}
	d^{G}
	=
	-3\alpha_s m_t
	\left(
	\frac{g_s}{4\pi}
	\right)^3
	\frac{\mathrm{Im}\!\left(\Gamma_t^{12}\right)}
	{m_{\tilde g}^{3}}
	(z_1-z_2)
	H(z_1,z_2,z_t),
	\label{eq:dG}
\end{equation}

where

\begin{equation}
	z_{\alpha}
	=
	\left(
	\frac{M_{\tilde t_{\alpha}}}
	{m_{\tilde g}}
	\right)^2,
	\qquad
	z_t
	=
	\left(
	\frac{m_t}
	{m_{\tilde g}}
	\right)^2,
	\label{eq:z_defs}
\end{equation}

and

\begin{equation}
	\Gamma_t^{12}
	=e^{-i\xi_3}
	D_{t22}
	D_{t12}^{*},
	\label{eq:Gamma12}
\end{equation}

where $D_t$ is the matrix that diagonalizes the stop mass-squared matrix, and $H(z_1,z_2,z_t)$ is the loop function defined in Ref.~\cite{ Dai}.

\end{document}